\documentclass[11pt]{article}

\newcommand\half{{\scriptstyle{\frac{1}{2}}}}
\def\parag{\hfil\break} 
\def\kikezd{\parag\underbar}

\input epsf
\begin{document}

\setlength{\baselineskip}{16pt}

\title{Celestial mechanics, from high-school to Newton's Principia}

\author{
P.~A.~Horv\'athy
\\
Laboratoire de Math\'ematiques et de Physique Th\'eorique\\
Universit\'e de Tours\\
Parc de Grandmont\\
F-37 200 TOURS (France)
\\
e-mail~: horvathy@univ-tours.fr
}

\date{\today}

\maketitle

\begin{abstract}
A high-school exercise is used to get an insight into
planetary motion.
\end{abstract}


\goodbreak

\section{Introduction: some history} 

Science contests for high-school students have 
a hundred-years-old tradition in Hungary \cite{Eotvos, KunRad}. 
The first one, open to students who just completed their high-school
cursus, was organized in 1894 to mark the 
appointment of L. E\"otv\"os as Minister of Education. 
This was also the year when the {\it High-School Journal 
 for Mathematics and Physics} was first published. This Journal
sets selected problems each month, and the best solutions
are published under the name of the student who solved it.
 
The early prize-winners of the E\"otv\"os contest include
such  outstanding future scientists as L. Fej\'er (who became famous
in mathematical analysis), 
Th. von K\'arm\'an (who made important contributions to hydro
and aerodynamics),
 A. Haar (remembered for his invariant measure on group manifolds)
 \dots  followed by many others. 

In 1916, in the middle of World War I,
a physics contest was held for the first time -- 
and L. Szil\'ard (who later, with E. Fermi, patented the nuclear reactor)
came second. E. Wigner (who got the Nobel prize for
using group theory in atomic physics),
and J. von Neumann (who built the first computer 
and contributed to many fields,
ranging from game theory to quantum mechanics)
only missed the contest
because of the turmoil following the disaster of World War~I.

In 1925 Edward Teller
not only got the first prize in physics, but also
shared the mathematics prize with Laszlo Tisza.
 They became friends and their friendship lasted until Teller's 
 recent death.
Soon Tisza (then a mathematics student in G\"ottingen)
changed to physics under the influence of Born, and
published his very first paper jointly with Teller (who was preparing
his Ph. D.  in Leipzig with Heisenberg) on molecular spectrum. 
Later Tisza introduced the two-fluid model of superfluidity, 
further developped by Landau. At 96, he is still active at MIT.

In the late twenties,
von Neumann and Szil\'ard even suggested that the university
examinations could be replaced by such a contest \cite{KunRad}.

 The tradition still continues, and contributes to forming  future
generations of scientists \cite{Eotvos}. 
The contests have been
internationalized with the {\it International Olympiads
for High-School Students}: first in Mathematics, and, since 1967,
also in Physics.

The E\"otv\"os contest lasts 5 hours, and the use of {\it all} documents or tools is 
allowed. Three problems that
require imagination and creative
thinking rather than lexical knowledge
are asked. They can, in some cases, lead to genuine research. 
A famous example is that of A. Cs\'asz\'ar (a distinguished
topologist), who, as
a young assistent, was called to survey the contest. While the 
high-school students were working, Cs\'asz\'ar figured out
a generalization of the geometric
problem given that year,
and later published a paper on his findings.

\section{A problem of spacecraft landing}

Some of the physics problems deserve further thinking also.
In 1969  -- the year of the  Moon landing, --  for example,
an exercise asked the following.
{\it A spacecraft moves on a circular trajectory of radius $r=4000$ km
around a planet of radius $R=3600$ km.
Then, for a short time, the 
rocket engins (directed oppositely to its motion), 
are switched on. This puts the spacecraft onto an
elliptic trajectory, which touches the planet's surface at the opposite
 point of the trajectory. Which proportion of its kinetic energy
went lost~?}

The problem can be solved using elementary 
methods  alone \footnote{As the impact speed does not vanish,
the problem is 
obviously {\it not} a realistic model of Moon landing -- but this has 
not been its aim, neither.}. It is, however, instructive
to describe it also using different, more sophisticated, methods,
which provide us with an insight into the problem of planetary
motion.
\goodbreak

From a physical point of view, we have the following situation.
As the engine works for a very short time, the position of the 
spacecraft does not change considerably. Owing to its loss of 
velocity, the gravitational attraction wins the race and pulls 
the circular 
trajectory more tight~: the trajectory
becomes an ellipse with major axis $2a=r+R$. Our task amounts to
comparing the kinetic energies of the two types of motions in the same
(aphelium) point. 
\goodbreak
\vskip2mm
\centerline{\epsfxsize=6.3truecm\epsfbox{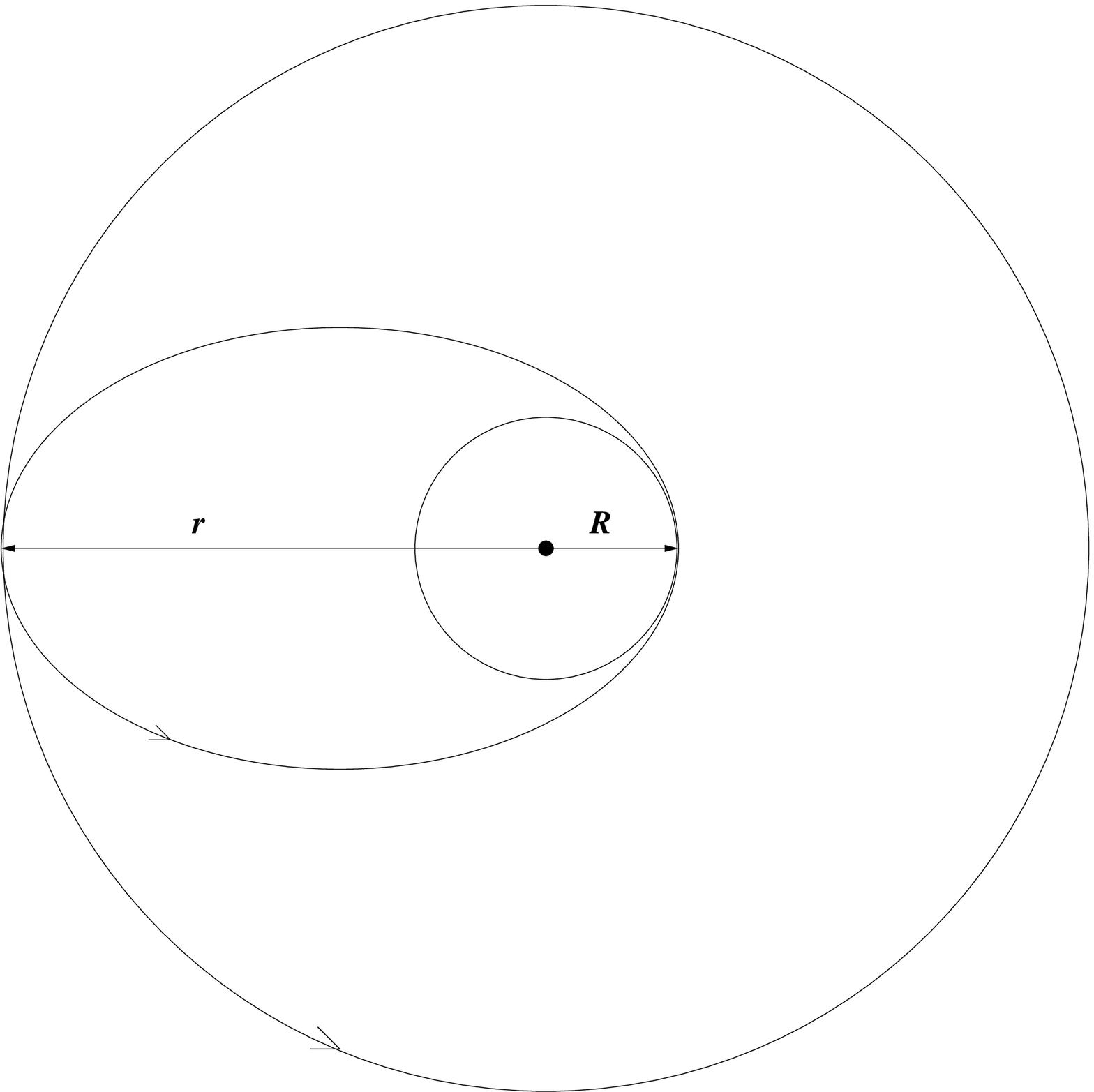}}
\vskip-0.2truecm

\kikezd{Figure 1:}\hskip2mm
{\it Reducing the velocity puts the spacecraft onto an
elliptic trajectory. 
The relative loss of energy is the numerical excentricity of the
new orbit.}
\vskip2mm
\goodbreak

The clue is to establish the following statement:
{\it  The ratio of the [aphelium] kinetic energies of the two types of 
motion is the perihelium-distance divided by the major semi-axis
of the elliptical motion},
\begin{equation}
    \delta=\frac{E_{a}^{kin}}{E_{0}^{kin}}=
    \frac{r_{p}}{a}=\frac{2R}{r+R}.
    \label{arany}
\end{equation}
Then the answer to the originally asked question follows
at once:
\begin{equation}
    \epsilon=\frac{E_{0}^{kin}-E_{a}^{kin}}{E_{0}^{kin}}
    =1-\delta=\frac{r-R}{r+R}.
    \label{1/19}
\end{equation}
Now, as
$ 
e=\half(R+r)-R=\half(r-R)
$ 
is the excentricity, (\ref{1/19}) is indeed $\epsilon =e/a$~:
{\it the relative loss of energy is the numerical excentricity of the
new orbit}, i. e., the measure of the ``flattening" of
the circle. 
Let us observe that the answer only depends on the geometric
dimensions. With the given numerical data, we find
$\epsilon=1/19$.

Below we present several proofs of 
(\ref{arany}), ordered following their ever-increasing difficulty
and required background knowledge.
\goodbreak

\section{Demonstrations}


\kikezd{Proof I: using Kepler's laws alone}.

The first proof is elementary.
According to the laws of circular motion,
\begin{equation}
    \frac{v_{0}^2}{r}=\frac{fM}{r^2}
    \label{korero}
\end{equation}
where $f$ is Newton's constant. The [square of the] period is hence
\begin{equation}
     T_{0}^2=\big(\frac{2\pi r}{v_{0}}\big)^2=4\pi^2\frac{r^3}{fM}.
\end{equation}

 This circular motion has kinetic energy
\begin{equation}
    E_{0}^{kin}=\frac{mv_{0}^2}{2}=\frac{fMm}{2r}.
    \label{korenerg}
\end{equation}
Application of Kepler's third law to the elliptic motion yields
the period of this latter,
\begin{eqnarray}
\frac{T^2}{T_{0}^2}=\frac{(R+r)^3}{(2r)^3}
&\Longrightarrow
&T=\frac{2\pi}{\sqrt{fM}}\left(\frac{R+r}{2}\right)^{3/2}.
\end{eqnarray}

The area of the ellipse is
$\pi a b$, where $b$ is the minor semi-axis.  
$b^2=\sqrt{a^2-e^2}=\sqrt{rR}$, the area is hence 
$\pi(R+r)\sqrt{rR}/2$.
The areal velocity, which is constant by Kepler's second law,
is therefore
\begin{equation}
    \nu=\frac{\pi ab}{T}=\sqrt{\frac{fMrR}{2(R+r)}}.
\end{equation}
At the aphelium $\nu=\half v_{a}r$, so that after slowing down,
the velocity is
\begin{equation}
    v_{a}=\sqrt{\frac{2fMR}{r(R+r)}}.
\end{equation}
The corresponding kinetic energy is then
\begin{equation}
    E_{a}^{kin}=\frac{mv_{a}^2}{2}=\frac{fMm}{r}\frac{R}{R+r}
    =E_{0}^{kin}\frac{2R}{R+r}
    \label{aphenerg}
\end{equation}
which implies (\ref{arany}).

\goodbreak
\kikezd{Proof II: Using the conservation of the energy
and of the angular momentum}.

Denoting the aphelium and the perihelium velocities by
$v_{a}$ and $v_{p}$, 
the conservation of the energy and of the angular momentum
(divided by $m$) requires that
\begin{eqnarray}
    \frac{v_{a}^2}{2}-\frac{fM}{r}=\frac{v_{p}^2}{2}-\frac{fM}{R},
    \label{ellenerg}
    \\[5pt]
    \frac{v_{a}\cdot r}{2}=\frac{v_{p}\cdot R}{2}.
    \label{ellimp}
\end{eqnarray}

Eliminating the perihelium velocity yields once again the kinetic energy
 (\ref{aphenerg}).
\goodbreak

\kikezd{Proof III: Using the formula of the total energy
of planetary motion}.\label{III}

An important property of planetary motion \cite{Goldstein, Landau, Cordani} is that the total energy 
only depends on the major axis, according to
\begin{eqnarray}
    E^{tot}=-\frac{fMm}{2a}.
    \label{totenerg}
\end{eqnarray}
Then it follows from the energy conservation
that, in any point of the trajectory,
the velocity satisfies
\begin{equation}
    v^2=fM\left(\frac{2}{r}-\frac{1}{a}\right).
    \label{lemma}
\end{equation}

For the circular motion, $a=r$, and for the elliptic $a=(r+R)/2$, 
respectively. Plugging this into
(\ref{lemma}), yields
 (\ref{korenerg}) and (\ref{aphenerg}), respectively.
 
Even more simply, observing that the change of the total
energy is in fact that of the kinetic energy since the 
potential energy is unchanged, using
 (\ref{totenerg}) we have
\begin{equation}
    \Delta E^{kin}=\Delta E^{tot}= 
    fMm\left(\frac{1}{r+R}-\frac{1}{2r}\right)=
    \big(\frac{fMm}{2r}\big)\frac{r-R}{r+R}.
    \label{energiavalt}
\end{equation}
Writing here by (\ref{korenerg}) $H_{0}^{kin}$ in place of
 $fM/2r$ yields (\ref{1/19}) directly.

It is worth noting that, at a point $r$ of the trajectory,
the ratio of the kinetic and the potential energies is,
by (\ref{lemma}), 
$
E^{kin}/E^{pot}=r/2a-1.
$
We have therefore
\begin{eqnarray}
E_{0}^{kin}=-\half E^{pot},
\qquad\hbox{resp.}\qquad
E_{a}^{kin}=-\frac{R}{r+R} E^{pot};
\label{kinpot}
\end{eqnarray}
which yields again (\ref{arany}).

Furthermore, while the total energy only depends on
the major semi-axis, this is not so for the parts taken individually by the
kinetic and the potential energies. According to
(\ref{kinpot}) we have indeed
\begin{eqnarray} 
E_{0}^{kin}=-E_{0}^{tot},
\qquad
E_{a}^{kin}=-\frac{R}{r}E_{a}^{tot}.
\label{kintoten}
\end{eqnarray}
\goodbreak

\kikezd{Proof IV: From the radial equation}

The result can also be obtained from studying radial
motion\footnote{This was suggested to me by J. K\"urti.}. 
For an arbitrary central potential $V$, the problem can be reduced
to one-dimensional motion with effective potential
$V_{eff}=V+\frac{L^2}{2mr^2}$ where
$L$ is the total angular momentum, cf. \cite{Goldstein} p. 75. 
The radial velocity is therefore given by
\begin{equation}
    \dot{r}=\sqrt{\frac{2}{m}\left(E-V-\frac{L^2}{2mr^2}\right)}
    \label{radvel}
\end{equation}
where $E=E^{tot}$ is the total energy. 
In the extremal points $r_{a}$ and $r_{p}$ 
the radial velocity vanishes, so that, for $V=-fMm/r$,
$$
-\frac{fmM}{r_{a}}+\frac{L^2}{2mr_{a}^2}
=E
-\frac{fmM}{r_{p}}+\frac{L^2}{2mr_{p}^2}.
$$
Thus
$$ 
    \left(\frac{L}{m}\right)^2=\frac{2}{
    \displaystyle{\frac{1}{r_{a}}+\frac{1}{r_{p}}}}\,fM.
    \label{angmomsq}
$$ 
But the aphelium velocity is $v_{a}=L/mr_{a}$ so that
\begin{equation}
    v_{a}^2=\frac{fM}{r_{a}^2}\cdot
    \frac{2}{\displaystyle{\frac{1}{r_{a}}+\frac{1}{r_{p}}}}
    \label{velsq}
\end{equation}
For a circular motion at $r=r_{a}$,
$fM/r_{a}=v_{0}^2$, and therefore
\begin{equation}
 \frac{v_{a}^2}{v_{0}^2} =\frac{1}{r_{a}}\cdot
 \frac{2}{\displaystyle{\frac{1}{r_{a}}+\frac{1}{r_{p}}}},
\end{equation}
consistently with our previous result.

\kikezd{Proof V: Relation to Kepler's third law}.

Kepler's third law is related to the behaviour of
the system with respect to scaling \cite{Landau}~:
if some trajectory is dilated from the focus by 
 $\lambda$, and the time is dilated by $\lambda^{3/2}$, 
\begin{equation}
{\bf r}\to {\bf r}'=\lambda\,{\bf r},
\qquad
t\to t'=\lambda^{3/2}\,t,
\end{equation}
yields again a possible trajectory. 
In those points which correspond to each other, both the
kinetic and the potential energies [and hence also the total energy]
are related as the inverse ratio of the geometric dimensions,
\begin{equation}
     \frac{E'}{E}=\lambda^{-1}.
     \label{KIII}
\end{equation}

Let us now retract our original circular motion so that its
radius equals to the major semi-axis of our elliptic motion above,
i.e., consider the dilation by $\lambda=\half(r+R)/r$.
By (\ref{KIII}) the total energy [and consistently with 
(\ref{totenerg})] is
$$
{\tilde{E}}_{0}^{tot}=\frac{2r}{r+R}E_{0}^{tot}.
$$  
This is, however, the same as the total energy of the elliptic motion,
$\tilde{E}_{0}^{tot}=E_{a}^{tot}$, 
since the major semi-axes are equal. Hence once again
$$ 
E_{a}^{tot}=\frac{2r}{r+R}E_{0}^{tot}.
$$
Then the result follows from (\ref{kintoten}).

Let us stress that Kepler's third law did not suffice alone;
we also needed the statement about the total energy.
\goodbreak

\kikezd{Proof VI: Using the Frenet formul{\ae}}.

It is worth describing the motion using the moving local
frame introduced by Fr\'enet \cite{Frenet}. Then, for a trajectory 
of arbitrary shape, the normal component of the acceleration is
 ${v^2}/{\rho}$ where $\rho$ is the radius of curvature
 i. e., the radius of the osculating circle \cite{Frenet}.
 In an extremal point of the ellipse
 the accelaration is normal, and points toward the focus. Hence
\begin{equation}
   m \frac{v^2}{\rho}=\hbox{Force}
    \label{altalanos}
\end{equation}
which generalizes the formula  (\ref{korero}) of circular motion.
For the circle  $\rho=r$, so that
\begin{equation}
    \frac{v_{0}^2}{r}=\frac{v_{a}^2}{\rho}
    \qquad\Longrightarrow\qquad
    \frac{E_{a}^{kin}}{E_{0}^{kin}}=\frac{\rho}{r},
    \label{azonnal}
\end{equation}
since the force is the same for both problems.
We have hence proved:
{\it The ratio of the kinetic energies is 
identical to that of the radii of curvature}.
In the extremal points of the ellipse,
$$
\rho=\frac{b^2}{a}=\frac{2rR}{r+R},
$$
which implies again
 (\ref{arany}). This confirms our intuition:
decreasing the velocity increases the curvature.
Using the explicit form,
${fMm}/{r^2}$, of the force,
(\ref{altalanos}) would allow us to calculate the
velocity as
\begin{equation}
    \frac{v_{a}^2}{2}=(\frac{fM}{2r})\cdot\frac{\rho}{r}
    =(E_{0}/m)\cdot\frac{\rho}{r}.
    \label{19}
\end{equation}
This is, however, not necessary for us: it was enough to know
the geometric dimensions of the trajectory.
\goodbreak

\kikezd{Proof VII: Using the ``Runge-Lenz'' vector}. 

A proof analogouos to that in II is obtained if we use
the so called ``Runge-Lenz'' vector \cite{Goldstein, RungeLenz,
Landau, Cordani}
\begin{equation}
    {\bf K}=m{\bf v}\times{\bf L}-fMm\,\hat{\bf r}
    \label{Rungelenz}
\end{equation}
where  ${\bf L}={\bf r}\times{\bf v}$ is the conserved
angular momentum;  $\hat{\bf r}$ denotes  the unit vector
carried by the radius vector drawn form the Earth's center 
to the spacecraft's position.

\goodbreak
\vskip2mm
\centerline{\epsfxsize=7.5truecm\epsfbox{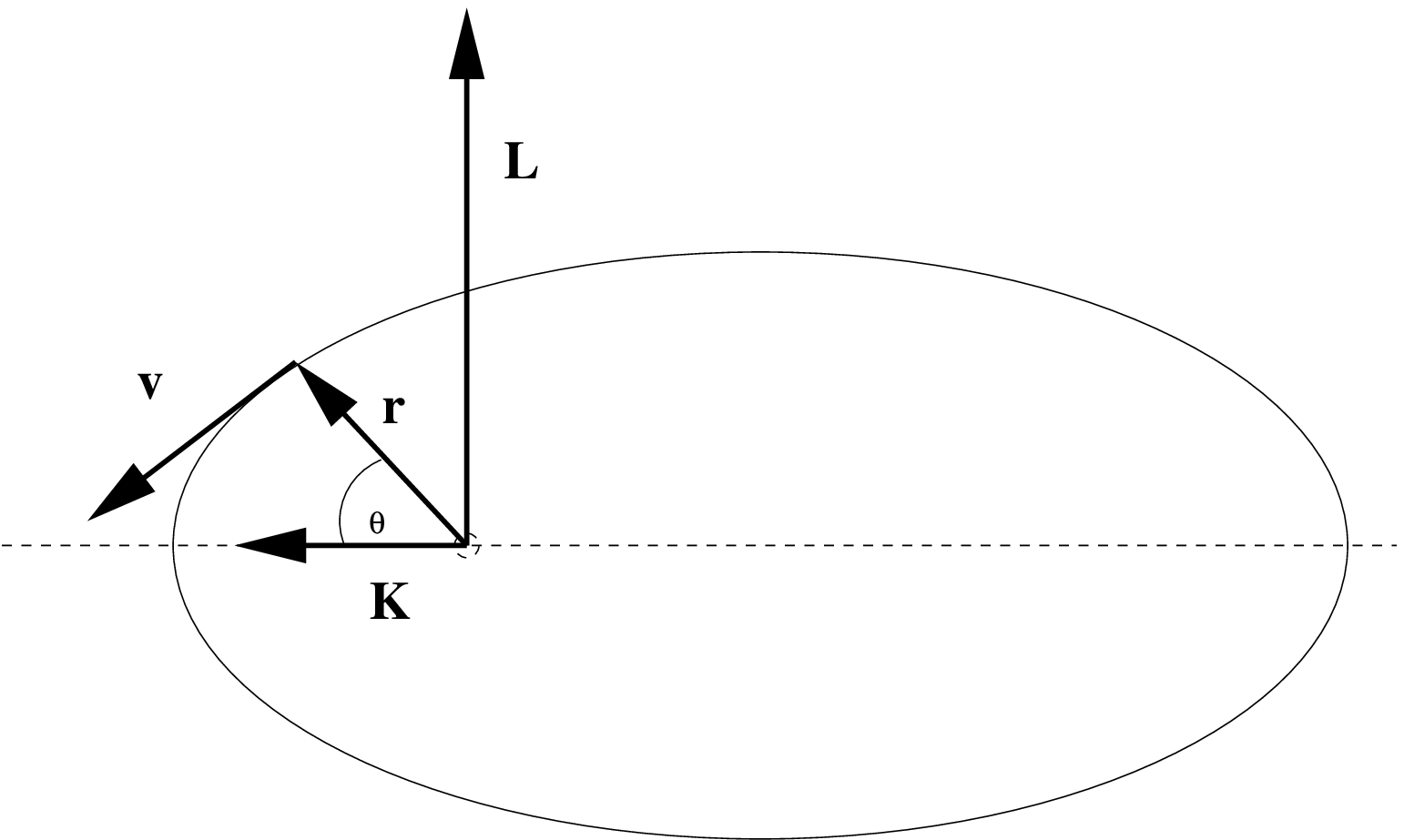}}
\vskip-0.4truecm

\kikezd{Figure 2:}\hskip2mm
{\it The conserved Runge-Lenz vector is directed from the
Earth' center towards the perihelium position.}
\vskip2mm
\goodbreak

Differentiating 
${\bf K}$-t with respect to time shows,
using
$$
\dot{\hat{\bf r}}=\frac{\bf v}{r}-\frac{({\bf r}\cdot{\bf 
v})}{r^3}{\bf r}
=\frac{1}{r^3}\left({\bf r}\times({\bf v}\times{\bf r})\right)=
-\frac{1}{r^3}{\bf r}\times({\bf L}/m)
$$
and the equation of motion, that 
${\bf K}$ is a constant of the motion.
The scalar product of  
${\bf K}$ with ${\bf L}$ vanishes, so that 
 ${\bf K}$ lies in the plane of the motion; it points from the
 focus to the perihelium point~: ${\bf K}=K\hat{\bf e}$.
Multiplying (\ref{Rungelenz}) with ${\bf r}$
 yields the trajectory \cite{Goldstein, Landau}  as
\begin{equation}
    r=\frac{p}{1+\epsilon\cos\theta}
    \qquad
    p=\frac{L^2}{fMm^2}
    \qquad
    \epsilon=\frac{K}{fMm}
    \label{ellegy}
\end{equation}
where $\theta$ is the angle between ${\bf K}$ and ${\bf r}$.
(\ref{ellegy}) defines a conic with
parameter $p$ and numerical excentricity $\epsilon$.

Returning to our initial problem, let us observe that in the extremal points
\begin{equation}
    {\bf K}=mv_{p}L\hat{\bf e}-fMm\hat{\bf e}
    =
    -mv_{a}L\hat{\bf e}+fMm\hat{\bf e},
    \label{extrRL}
\end{equation}
where $\hat{\bf e}$ is the unit vector directed
from the center to the  perihelium.
The length of ${\bf L}$ is clearly
$L=mv_{p}r_{p}=mv_{a}r_{a}$ [cf. (\ref{ellimp})];  eliminating the 
perihelium velocity,
\begin{equation}
    \frac{v_{a}^2}{2}=fM\frac{r_{p}}{r_{a}(r_{p}+r_{a})}.
\end{equation}

For circular motion $r_{a}=r_{p}=r$ yielding (\ref{korenerg});
for our elliptic motion 
$r_{p}=R$, $r_{a}=r$ which provides us again with 
(\ref{aphenerg}), the kinetic energy in the aphelium.
Squaring (\ref{extrRL}) yields furthermore
\begin{equation}
K^2=f^2M^2m^2+2E^{tot}L^2.
\label{casimir}
\end{equation}
Hence $K=fMm\epsilon$ which, together with (\ref{extrRL}) yields
\begin{equation}
    v_{a}^2=\frac{fM}{r}(1-\epsilon).
    \label{excenten}
\end{equation}
Writing  $2E_{0}^{kin}$ for $fMm/r$ provides us again with 
(\ref{arany}) or (\ref{1/19}).
\goodbreak

\kikezd{Proof VIII. Using the hodograph}

Drawing the instantaneous velocity vector from a fixed point $O$ of
``velocity space'' yields the {\it hodograph}.
For planetary motion this is a circle \cite{Goldstein, Cordani}.
The simplest proof of this statement is obtained if we multiply 
the angular momentum vectorially with
the Runge-Lenz vector  \cite{Goldstein}. Developping the
double vector product and using
${\bf L}\cdot{\bf v}=0$ yields
$$
{\bf L}\times{\bf K}=L^2m{\bf v}-fMmL{\bf u}
$$
where ${\bf u}$ is the unit vector
obtained by rotating $\hat{\bf r}$ around the direction
of ${\bf L}$ by  $90{}^o$ degrees in the counter-clockwise direction.
As ${\bf L}\times{\bf K}=-LK{\bf j}$ where ${\bf j}$ is the unit vector
directed along the $y$ axis of the coordinate plane.
Writing here $K=fMm\epsilon$ the velocity vector is expressed as
\begin{equation}
    {\bf v}=\frac{fM}{L}\,\Big({\bf u}
    -\epsilon\,{\bf j}\Big).
    \label{hodo}
\end{equation}

As the unit vector $\hat{\bf r}$ 
turns around during the motion, so does also ${\bf u}$
(advanced with ($90{}^o$ degrees). The first term in
(\ref{hodo}) describes hence a circle of radius $fM/L$,
whose center has been translated to $C$, situated on the $y$ axis at
distance $-K/L=fM\epsilon/L$ below $O$.
For a circular trajectory $C=O$ and the hodograph becomes a circle
around the origin with radius $v_{0}^2=fM/r$.

\goodbreak
\centerline{\epsfxsize=6.5truecm\epsfbox{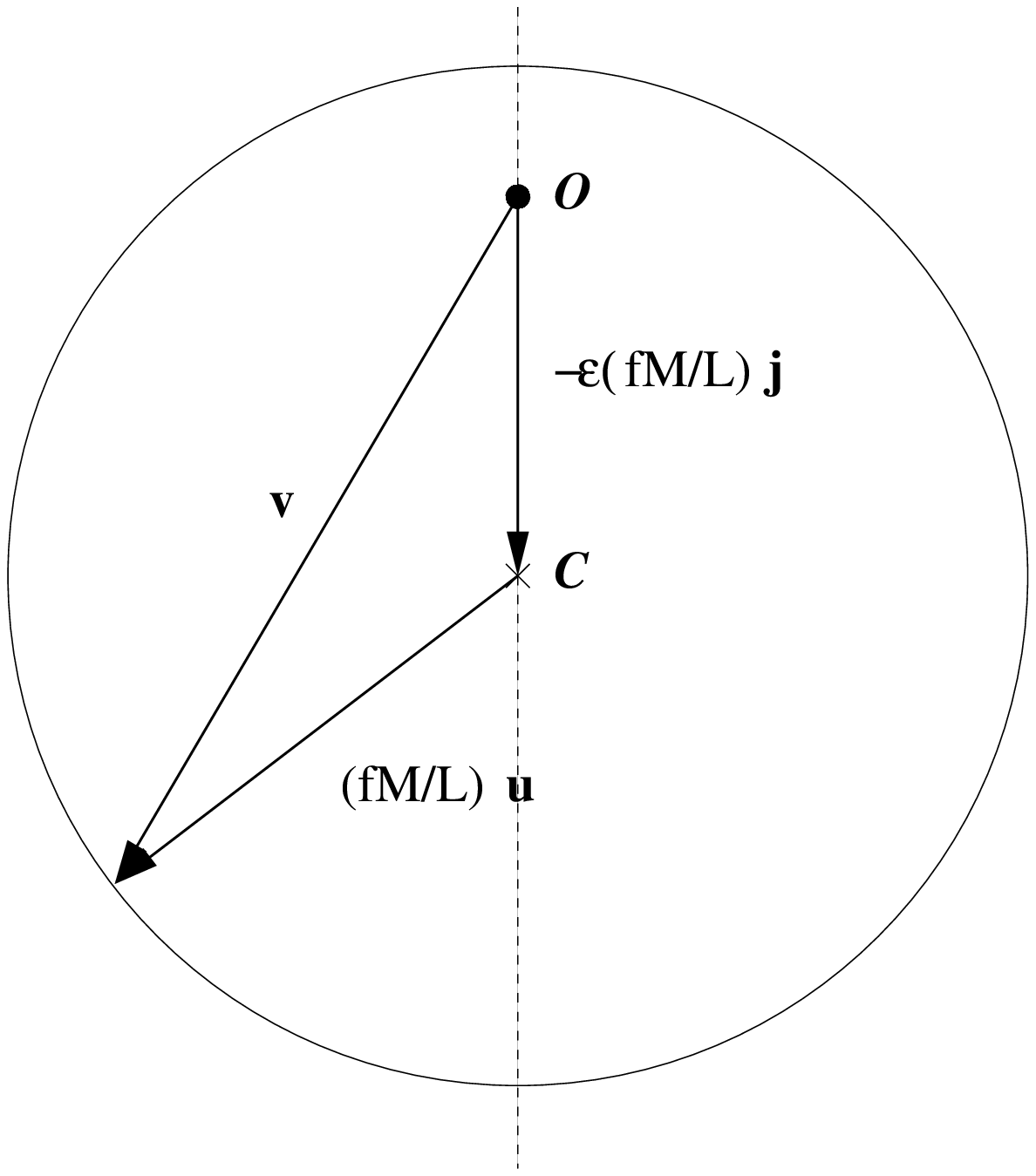}}
\vskip-0.6truecm

\kikezd{Figure 3:}
{\it The velocity vector of planetary motion describes a circle
of radius $fM/L$ and center at $C$,
situated on the $y$ axis at distance ($\overline{OC}=K/L=\epsilon fM/L$)
in ``velocity space''. The aphelium point corresponds to the shortest
distance from $O$, represented as the top of the circle.
}
\vskip2mm
\goodbreak

The velocity is the largest at the bottom of the circle, which
corresponds to the perihelium point. The smallest velocity is obtained 
in turn in the aphelium, which is the top of the circle.
Then ${\bf u}$ points vertically upwards,
 ${\bf u}_{a}={\bf j}$. 
Then the length of this smallest velocity is plainly
$$
\hbox{smallest velocity is}\quad = \quad
\hbox{(radius) -- (distance}\; \overline{OC}),
$$
which yields (\ref{excenten}). Alternatively, we can write
$L=v_{a}r_{a}$ in (\ref{hodo}).

\kikezd{Proof IX. Maxwell's generalization}

It is worth noting that our problem here can be readily generalized.
J. C. Maxwell 
has in fact shown in a lovely little booklet
\cite{Maxwell} that, in an {\it arbitrary} point $r$
of the trajectory, the velocity is
\begin{equation}
    v^2=\frac{4\pi^2a^2}{T^2}\left(\frac{2a}{r}-1\right).
    \label{altseb}
\end{equation}

Maxwell' proof only uses Kepler's second law 
and some geometric properties of the ellipses.
Let in fact the Sun be in one of the foci, $S$, 
let $H$ be the other focus and
 draw perpendiculars $SY$ and $HZ$ from the foci
to the tangent to the ellipse at a point $P$, see Fig. 4.
\vskip2mm
\centerline{\epsfxsize=9.5truecm\epsfbox{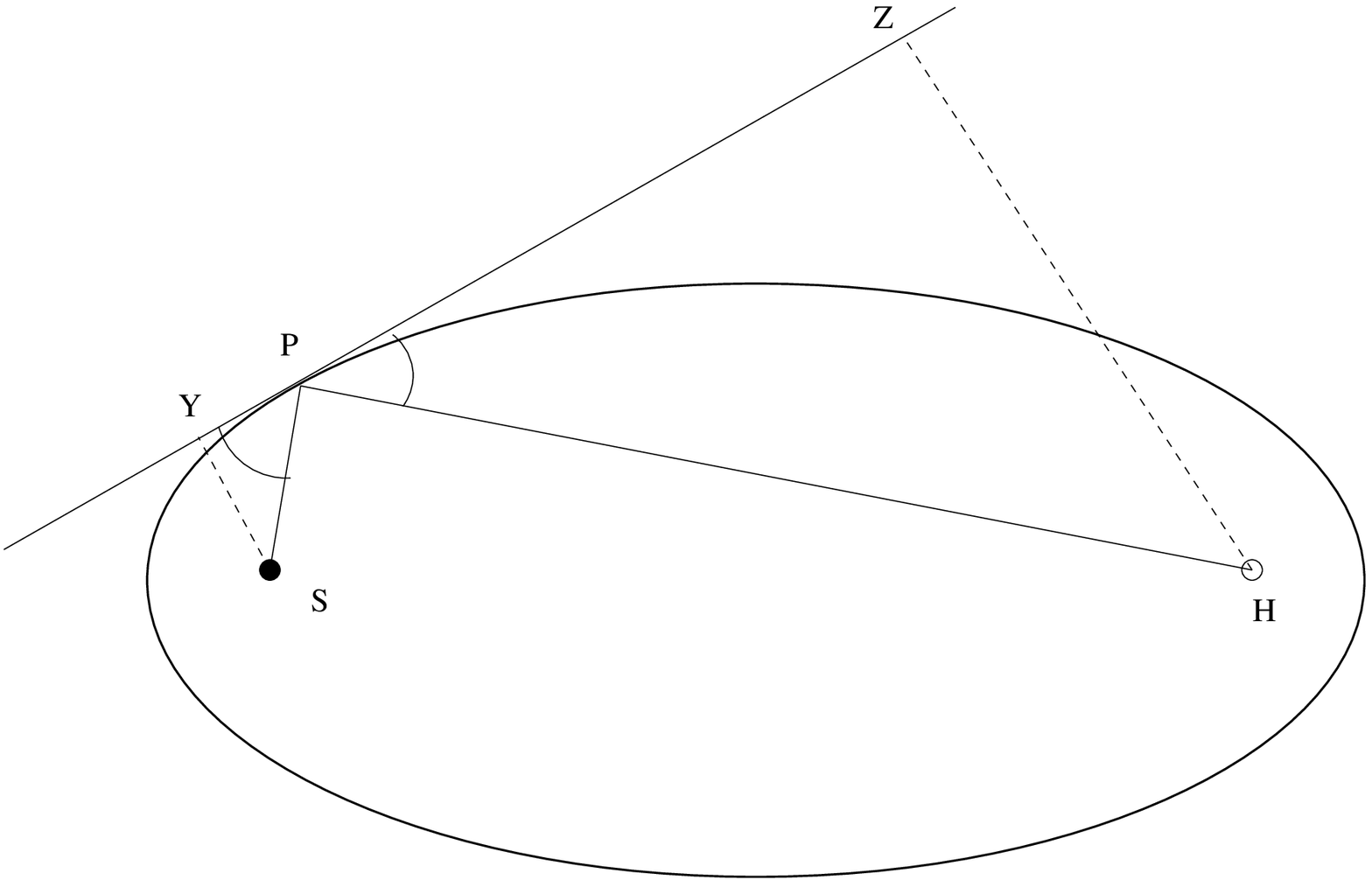}}
\vskip-0.6truecm

\kikezd{Figure 4:}
{\it Maxwell's geometric derivation.}
\vskip2mm
\goodbreak
By Kepler's second law, the areal velocity is constant and
given by twice the area of the ellipse, $\pi ab$, divided
by the revolution time, $T$,
\begin{equation}
\overline{SY}\cdot v=\frac{2\pi ab}{T}
\qquad\Longrightarrow\qquad
v=\frac{2\pi ab}{T} \frac{1}{\overline{SY}},
\label{velocity}
\end{equation}

On the other hand, a known geometric property of
the ellipse says that the half minor axis, $b$, is the mean of 
 $\overline{SY}$ and $\overline{HZ}$,
$$
 \overline{SY}\cdot\overline{HZ}=b^2.
$$

Finally,  the ``optical property'' [known to Kepler] says 
that all light rays emitted from one focus are 
reflected by ellipse to the other focus. It follows that
the angles $HPZ$ and $SPY$ are equal; 
 the triangles $SPY$ and $HPS$ are therefore similar. Hence
$$
\overline{SP}:\overline{PH}=\overline{SY}:\overline{ZH}.
$$
Multiplication of the two relations yields
$$
\overline{SY}^2=b^2\,\frac{\overline{SP}}{\overline{PH}}
=b^2\frac{r}{2a-r},
$$
since $\overline{SP}+\overline{PH}=2a$, the major axis
 of an ellipse.
Inserting $\overline{SY}^2$ into the square of (\ref{velocity})
yields Maxwell's formula (\ref{altseb}).
 
For a circular orbit in particular,
$v_{0}^2=4\pi^2r^2/T_{0}^2$. 
(In general, i. e. with the exception
of the extremal points, the velocities
${\bf v}_{0}$ and ${\bf v}$ have different directions.)
Hence
$$
\frac{v^2}{v_{0}^2}=\frac{a^2}{r^2}\frac{T_{0}^2}{T^2}
\frac{2a-r}{r}.
$$
Using here
Kepler's III law, we end up with the following simple expression:
\begin{equation}
 \delta=\frac{v^2}{v_{0}^2}=\frac{2a-r}{a}.
 \label{altarany}
\end{equation}
In words: {\it At any point of the orbit, the square of the velocity
to that of the circular motion through the same point is
as the distance from the other focus to the semi major axis}~!
For $r=r_{a}$ we recover in particular our previous 
result (\ref{arany}).

\goodbreak
\kikezd{Proof X: From Newton's Principia}

Last but not least, let us point out that the problem discussed 
here has actually been solved by Newton over 300 years ago~!
Proposition 16. Theorem 8.  Corollary 3. of 
his ``Principia'' \cite{Princip} says in fact that 
``{\it the velocity in a conic, at the greatest or least distance 
from the focus, is to the velocity with which the body would move in a 
circle, at the same distance from the center, as the
square root of the principal latus rectum is to the square root of 
twice that distance.}'' Translated into our notations,
\begin{equation}
    v_{a}^2 : v_{0}^2=\ell : 2r.
    \label{Newton}
\end{equation} 
But the half the ``principal latus rectum'' is the harmonic mean
of the periapsis and the apoapsis
of the ellipse  \cite{Coxeter} (and also the parameter $p=b^2/a$),
$$ 
\frac{2}{\ell}=\frac{1}{2}\left(\frac{1}{r_{p}}+\frac{1}{r_{a}}\right)
\quad \Longrightarrow\quad
\ell=\frac{4rR}{r+R}.
$$
Inserting into
(\ref{Newton}) the result (\ref{arany}) is recovered once again.

Newton's original proof comes by a series of elementary geometric 
demonstrations using hardly more then the properties of similar
triangles \cite{Princip}. The reader is referred to
Newton's Principia for details.
Let us remark, though, that (\ref{altarany}) provides a quick
proof to Newton's statement (\ref{Newton}), since the prodcuct
of $r=r_{a}=a(1+\epsilon)$ and $r_{p}=a(1-\epsilon)$ 
is $r_{a}r_{p}=a^2(1-\epsilon^2)=a\ell/2$.

\section{Discussion}

Let us summarize our various approaches.
Our first proof only used Kepler's laws specific
for the planetary motion, and suits perfectly to
 a high-school student.
 The second and seventh proof is based on conservation laws;
the second uses that of the energy and the angular momentum,
 and the last the Runge-Lenz vector. This is early 19th century physics:
the vector (\ref{Rungelenz}) was in fact introduced by Laplace in 1799,
in his  {\it Trait\'e de M\'ecanique C\'eleste}
\cite{Goldstein, RungeLenz, Cordani}.

Proof IV is based on the radial equation.

II, using high-school knowledge only,
would clearly work for any conservative force,
while VII is related to the ``hidden'' symmetry of of the
Kepler problem. Although some of the results
could be obtained by freshmen, this approach 
is not in general taught at the university. It leads
to a group theoretical treatment of planetary motion \cite{Cordani}. 
For example, ${\bf L}\cdot{\bf K}=0$ and 
(\ref{casimir}) are the classical counterparts of
the Casimir relations of the SO$(4)$ dynamical symmetry
used by Pauli to determine the spectrum of the
hydrogen atom \cite{Cordani}.

III and V are based on the formula (\ref{totenerg}) of the total 
energy, discussed by university textbooks \cite{Goldstein, Landau}.
V is linked to the scaling property
which yields in fact Kepler's third law \cite{Landau}. 

Proof V uses the general framework of co-moving
coordinates called the Fr\'enet formul{\ae} \cite{Frenet} 
(late 19th century),
which makes part of regular university courses on mechanics
and/or differential geometry.
It can be applied to any central force problem: 
the reader is invited to work 
out what happens, e, g., for a harmonic force
${\bf F}=-k{\bf r}$ (when the trajectories are again  ellipses.)

Proof VIII is based on the hodograph, also used
by Feynman in his geometric approach to planetary motion \cite{Feynman}.

The problem is readily generalized following Maxwell \cite{Maxwell}
(Proof IX),
 along the lines of Proof I and adding some knowledge of geometry.
 Remarkably, this general answer might have been found by Kepler
 in the early 17th century~!

Our final proof X uses a statement made by Newton in his
Principia, also obtained by elementary geometry.

Interestingly, the very first question of the
very first E\"otv\"os physics contest (that of Szilard in 1916),
was on planetary motion: {\it Let us consider
a miniaturized model of the solar system, where all distances are
scaled as $1:15\cdot 10^{10}$. How long would last a year~?}
The answer is that it would remain a year, since,
assuming that all densities remain constant, the Sun's
mass would also decrease as the cube of its dimension.
But this is precisely the condition,
as seen from the scaling argument in Proof V  \cite{Landau}.

\goodbreak


\end{document}